\documentclass[letterpaper]{article}
\usepackage{aaai}
\usepackage{times}
\usepackage{helvet}
\usepackage{courier}
 \usepackage{graphicx}
\usepackage{amsmath}
\usepackage{subfigure}

\RequirePackage{algorithm}
\RequirePackage{algorithmic}

\begin{document}
%
\title{Co-evolution of Selection and Influence in Social Networks }
\author{Yoon-Sik Cho \and Greg Ver Steeg \and Aram Galstyan\\
USC Information Sciences Institute\\
Marina del Rey, California 90292\\
}

\maketitle
\begin{abstract}
\begin{quote}
Many networks are  complex dynamical systems, where both attributes of nodes and topology of the network (link structure) can change with time. We propose a model of co-evolving networks where both node attributes and network structure evolve under mutual influence. Specifically, we consider a mixed membership stochastic blockmodel, where the probability of observing a link between two nodes depends on their current membership vectors, while those membership vectors themselves evolve in the presence of a link between the nodes. Thus, the network is shaped by the interaction of stochastic processes describing the nodes, while the processes themselves are influenced by the changing  network structure. We  derive an efficient variational inference procedure for our model, and validate the model on both synthetic and real--world data. 
\end{quote}
\end{abstract}

\section{Introduction}
The recent surge in online social media has made it possible to examine social networks at an unprecedented scale. Thus, it is important to have scalable approaches for modeling and understanding statistical and dynamical properties of such systems. Most real--world networks are inherently complex dynamical systems, where both attributes of the nodes and topology of the network can change with time.  Furthermore, those changes are often  intertwined with each other, providing complex feedback mechanisms between node and link dynamics. As an illustrative example of such an interplay in social networks, here we will focus on the processes of {\em selection} and {\em influence}. The former means that nodes tend to interact with similar nodes, whereas the latter asserts that  the evolution of a node's attributes are affected by its neighbors.


The problem of properly characterizing selection and influence has been a subject of extensive studies in sociology. 
For instance, \cite{Steglich07interuniversitycenter} suggested a continuous time agent--based model of network co--evolution. In this model, each agent is characterized by a certain utility function that depends on  the agent's individual attributes as well as his/her local neighborhood in the network. The agents evolve as  continuous--time Markovian processes which, at randomly chosen time points, select an action to maximize their utility. Despite its intuitive appeal, a serious shortcoming of this model is that it cannot handle missing data well, thus most of the attributes have to be fully observable. This was addressed in \cite{Fan2009} where a continuous Dynamic Bayesian approach was developed.

Continuous--time models have certain advantages when the network observations are infrequent and  well--separated in time. In situations where more fine-grained data is available, however,  discrete--time models are more suitable~\cite{Hanneke2009}. Here we suggest a discrete time dynamical network model that accounts for both selection and  influence. Our model model is based on Mixed Membership Blockmodel~\cite{MMSB}. MMSBs are an extension of stochastic block-models that have been studied extensively both in social sciences and in computer science~\cite{Holland1983,Goldenberg2010}. In a stochastic blockmodel each node is assigned to a block (or a role), and the pattern of interactions between different nodes depends only on their block assignment.  Many situations, however, are better described by multi--faceted interactions, where nodes can bear multiple latent roles that influence their relationships to others. MMSB accounts for such ``mixed" interactions, by allowing each node to have a probability distribution over roles, and by making the interactions role--dependent~\cite{MMSB}.

Our Co--evolving Mixed Membership Stochastic Blockmodel, or CMMSB,  provides a dynamic generalization of the mixed membership model by explicitly modeling the variation in the node membership vectors. Previously, a dynamic extension of the MMSB (dMMSB) was suggested in~\cite{dMMSB}.  In contrast to dMMSB, where the dynamics was imposed {\em externally}, our model assumes that the  membership evolution is driven by the interactions between the nodes through a parametrized {\em influence} mechanism. At the same time, the patterns of those interactions themselves change due to the evolution of the node memberships.

Another advantage of our model over dMMSB is that the latter models the aggregate dynamics, e.g.,  the {\em mean} of the logistic normal distribution from which the membership vectors are sampled. CMMSB, however, models each node's trajectory separately,  thus providing better flexibility for describing system dynamics.  Of course, more flexibility comes at a higher computational cost, as CMMSB tracks the trajectories of all nodes individually.  This additional cost, however,  can be well justified in scenarios  when the system as a whole is almost static (e.g., no shift in the mean membership vector), but different subsystems experience dynamic changes. One such scenario that deals with political polarization in the U.S. Senate is presented in our experimental results section.

\section{Co-evolving Mixed Membership Blockmodel}

Consider a set of $N$ nodes, each of which can have $K$ different roles, and let  $\vec{\pi}_p^t$ be the mixed membership vector of node $p$ at time $t$.  Let $Y_t$ be the network formed by those nodes at time $t$:  $Y_t(p,q)=1$ if the nodes $p$ and $q$ are connected at time $t$, and $Y_t(p,q)=0$ otherwise. Further,  let $Y_{0:T}=\{Y_0,Y_1,\dots,Y_T\}$ be a time sequence of such networks. The generative process that induces this sequence is described below.

\begin{itemize}
\item  For each node $p$ at time $t=0$, employ a logistic normal distribution over a simplex sample.~\footnote{We found that the logistic normal form of the membership vector  suggested in~\cite{dMMSB}  led  to more tractable equations compared to the Dirichlet distribution.} \\
$\pi_{p,k}^0 = \exp(\mu_{p,k}^0 -C(\vec{\mu}_p^0)), ~~~~~\vec{\mu}_p^0 \sim {\cal N} (\vec{\alpha}^0, A) $ \\
where $C(\vec{\mu}) = \log (\sum_k \exp(\mu_k))$ is a normalization constant, and $\vec{\alpha}^0, A$ are prior mean, and covariance matrix. 

 \item  For each node $p$ at time $t>0$, the mean of each normal distribution is updated due to {\em influence} from the neighbors at its previous step: \\
$ \vec{\alpha}^t_p =(1-\beta_p)\vec{\mu}_p^{t-1} + \beta_p\vec{\mu}_ {\mathcal{S}(p,t-1)}$\\
where $\vec{\mu}_ {\mathcal{S}(p,t-1)}$ is average of weighted membership vector $\vec{\mu}$-s of the nodes which node $p$ has met at time $t-1$\\
$\vec{\mu}_ {\mathcal{S}(p,t-1)} = \sum _{q} Y(p,q)w_{p\leftarrow q}^{t-1}\vec{\mu}_q$ \\
$\beta_p$ describes how easily the node $p$ is influenced by its neighbors.  The membership vector at time $t$ is \\
$\pi_{p,k}^t = \exp(\mu_{p,k}^t -C(\vec{\mu}_p^t)),~~~~~
\vec {\mu}_p^{t}  \sim {\cal N} 
(\vec{\alpha}^t_p,\Sigma_{\mu}  ) $\\ 
where the covariance $\Sigma_\mu$ accounts for noise in the evolution process.

\item For each pair of nodes $p$, $q$ at time $t$,  sample role indicator vectors from multinomial distributions: \\
   $\vec{z}^t_{p \rightarrow q} \sim Mult(\vec{z}|\vec{\pi}_p^t)$, $\vec{z}^t_{p \leftarrow q} \sim Mult(\vec{z}|\vec{\pi}_q^t)$\\ 
     Here $\vec{z}_{p \rightarrow q}$ is a unit indicator vector of dimension $K$, so that ${z}_{p \rightarrow q,k}=1$ means node $p$ undertakes role $k$ while interacting with $q$. 
    
     \item Sample a link between $p$ and $q$ as a Bernoulli trial: \\
   $Y_t(p,q) \sim Bernoulli (y| (1-\rho)\vec{z}^t_{p \rightarrow q}   B^t \vec{z}^t_{p \leftarrow q})    $  \\
  where $B$ is a $K\times K$ role--compatibility matrix, so that $B_{rs}^t$ describes the likelihood of interaction between two nodes in roles $r$ and $s$ at time $t$. When $B^t$ is diagonal, the only possible interactions are among the nodes in the same role. Also, $\rho$ is a parameter that accounts for the sparsity of the network. 
\end{itemize}

\noindent  
Thus, the coupling between dynamics of different nodes is introduced by allowing the role vector of a node to be influenced by the role vectors of its neighbors.To benefit from computational simplicity,  we updated $\vec{\pi}$ by changing its associated $\vec{\mu}$. This update of $\vec{\mu}$ is a linear combination of  $\vec{\mu}$ at its current state, and the values of its neighbors. The influence is measured by a node--specific parameter $\beta_p$, and $w_{p \leftarrow q} ^t$.
$\beta_p$ describes how easily the node $p$ is influenced by its neighbors: $\beta_p=0$ means it is not influenced at all, whereas $\beta_p=1$ means the behavior is solely determined by the neighbors. Conversely, $w_{p \leftarrow q} ^t$ reflects the influence that node $q$ exerts on  node $p$, so that larger  values correspond to more influence. 

\subsection{Inference and Learning}

Under the Co--Evolving MMSB, the joint probability of the data $Y_{0:T}$ and the latent variables $\{\vec{\mu}_{1:N}^t, \vec z_{p \rightarrow q}^t : p,q \in N, \vec z_{p \leftarrow q}^t :p,q \in N    \}$ can be written in the following factored form. To simplify the notation, we define $ \vec z_{p,q}^t$ as a pair of  $\vec z_{p \leftarrow q}^t$, and $ \vec z_{p \rightarrow q}^t$
\begin{eqnarray}
\label{eq:prob}
p(Y_{0:T}, \vec{\mu}_{1:N}^{0:T}, \vec Z_{\rightarrow}^{0:T}, \vec Z_{\leftarrow}^{0:T} | \vec{\alpha},A, B, \beta_p,w_{p\leftarrow q}^t, \Sigma_\mu) 
=  \nonumber~~~~\\
\prod_t \prod_{p,q} P(Y_t(p,q) | \vec z_{p,q}^t , B^t  ) 
 P(\vec z_{p,q}^t | \vec {\mu}_p^t, \vec {\mu}_q^t )  \nonumber \\  
\times P( \vec{\mu}_p^{t+1} |\vec{\mu}_p^t,\vec{\mu}_ {\mathcal{S}(p,t)}, Y_t, \beta_p  ) \prod_p P(\vec{\mu}_p^0 | \vec{\alpha}, A )
\end{eqnarray}
\noindent  
In Equation~\ref{eq:prob}, the term describing the dynamics of the membership vector is  defined as follows\footnote{For simplicity, we will assume $\Sigma_{\mu}$ is a diagonal matrix.}: 
\begin{eqnarray}
P(\vec{\mu} ^t _p | \vec{\mu} ^{t-1} _p, \vec{\mu} ^{t-1} _{\mathcal{S}(p,t)},\Sigma_\mu, Y_t, \beta_p) \nonumber ~~~~~~~~~~~~~~~~~~~~~~~~~~~~~~~~~~~~~\\
= f_G(\vec{\mu} ^t _p - f_{b} (\vec{\mu} ^{t-1} _p,\vec{\mu} ^{t-1} _{\mathcal{S}(p,t)} ),\Sigma_\mu) \label{update} \\
f _{G} (\vec{x}, \Sigma_\mu) = {1\over {(2{\pi})^{k/2}}  |\Sigma_\mu|^{1/2} }   e^{-{1\over 2} x^T {\Sigma_\mu}^{-1}x}   \label{gw}\\
f_{b} (\vec{\mu} ^{t-1} _p,\vec{\mu} ^{t-1} _{\mathcal{S}(p,t)} ) = (1-\beta ^{t-1} _p)\vec{\mu}^{t-1}_p + \beta ^{t-1}_p \vec{\mu}^{t-1}_{\mathcal{S}(p,t)}
\end{eqnarray} 

Performing exact inference and learning under this model is not feasible. Thus, one needs to resort to approximate techniques. Here we use a variational EM~\cite{Bernardo03thevariational,GMF} approach. The main idea behind variational methods is to posit a simpler distribution $q(X)$ over the latent variables with free (variational) parameters, and then fit those parameters so that the distribution is close to the true posterior in KL divergence.
\begin{equation}\label{KL}
D_{KL}(q||p) = \int_X q(X) \log {q(X)\over p(X,Y)} dX
\end{equation}
Here we introduce the following   factorized variational distribution:
\begin{eqnarray}
q(\vec{\mu} _{1:N} ^{0:T}, Z_{\rightarrow}^{0:T}, Z_{\leftarrow}^{0:T} | \vec{\gamma}_{1:N}^{0:T} , \Phi_{}, \Phi_{} )
= 
 \prod _{p,t} q_1(\vec{\mu}_p ^t | \vec{\gamma}_p^t , \Sigma_p^t ) \nonumber \\ 
\times  \prod_{p,q,t}(q_2(\vec{z}_{p \rightarrow q}^t | \vec {\phi} _{p \rightarrow q} ^t   ) q_2(\vec{z}_{p \leftarrow q}^t | \vec {\phi} _{p \leftarrow q} ^t   ))
\end{eqnarray} 
where $q_1$ is the normal distribution, and $q_2$ is the multinomial distribution, and $\vec{\gamma}_p^t , \Sigma_p^t ,\vec {\phi} _{p \rightarrow q} ^t , \vec {\phi} _{p \leftarrow q} ^t  $ are the variational parameters. Intuitively, $\phi^t_{p \rightarrow q,g}$ is the probability of node $p$ undertaking the role $g$ in an interaction with node $q$ at time $t$, and $\phi^t_{p \leftarrow q,h} $ is defined similarly. Note that in the E--step, we need to compute the expected value of $\log[\sum_k  \exp (\mu_k)]$ under the variational distribution, which is problematic. Toward this end, we introduce  $N$ additional variational parameters $\zeta$, and replace the expectation of the log by its upper bound induced from the first-order Taylor expansion:
\begin{equation}
\log[\sum \exp (\mu_k)]  \leq \log{\zeta} -1 + {1\over {\zeta}}\sum \exp (\mu_k)
\end{equation}

The variational EM algorithm works by iterating between the E--step of calculating the expectation value using the variational distribution, and the M--step of updating the model (hyper)parameters so that the data likelihood is locally maximized. The pseudo-code is shown in Algorithm~\ref{alg:example}, and the details of the calculations are discussed below.
\begin{algorithm}[tb]
   \caption{Variational EM}
   \label{alg:example}
\begin{algorithmic}
   \STATE {\bfseries Input:} data $Y_t(p,q)$, size $N, T, K$
   \STATE Initialize all  $\{\vec{\gamma}\}^t$, $\{\sigma\}^t$
   \STATE Start with an initial guess for the model parameters.    
   \REPEAT
   \REPEAT
   \FOR{$t=0$ {\bfseries to} $T$}
	\REPEAT
	\STATE Initialize $\phi_{p \rightarrow q}^t$,  $\phi_{p \leftarrow q}^t$ to $1\over K$ for all $g,h$
		\REPEAT
		\STATE Update all $\{\phi\}^t$  
		\UNTIL{convergence of $\{\phi\}^t$}
	\STATE Find $\{\sigma\}^t$, $\{\vec{\gamma}\}^t$
	\STATE Update all $\{\zeta\}^t$ 
	\UNTIL{convergence in time $t$ }

   \ENDFOR
   \UNTIL{convergence across all time steps}

   \STATE Update hyper parameters. 
   \UNTIL{convergence in hyper parameters}
\end{algorithmic}
\end{algorithm}

\subsection{Variational E-step}
In the variational E--step, we minimize the KL distance over the variational parameters. Taking the derivative of KL divergence with respect to each variational parameter and setting it to zero, we obtain a  set of equations that can be solved via iterative  or other numerical techniques. For instance, the variational parameters $(\vec{\phi}^t_{p \rightarrow q},\vec{\phi}^t_{p \leftarrow q} )$, corresponding to a pair of nodes $(p,q)$ at time $t$, can be found via the following iterative scheme: 
\begin{eqnarray}
\phi^t _{p \rightarrow q , g }
 {\propto}   \exp(\gamma_{p,g}^t) \nonumber~~~~~~~~~~~~~~~~~~~~~~~~~~~~~~~~~~~~~~~~~~~~~~~~~~~~~~~~~\\
 \times
\prod _h (B(g,h)^{Y_t(p,q)}(1-B(g,h))^{1-Y_t(p,q)} ) ^{\phi^t _{p \leftarrow q,h}} \label{uphi1}\\
\phi^t _{p \leftarrow q , h } 
  {\propto}  \exp(\gamma_{q,h}^t) \nonumber~~~~~~~~~~~~~~~~~~~~~~~~~~~~~~~~~~~~~~~~~~~~~~~~~~~~~~~~~\\
 \times  \prod _g (B(g,h)^{Y_t(p,q)}(1-B(g,h))^{1-Y_t(p,q)} ) ^{\phi^t _{p \rightarrow q,h}} \label{uphi2}
\end{eqnarray}
In the above equations, $\phi^t_{p \rightarrow q,g}$ and $\phi^t_{p \leftarrow q,h} $ are normalized after each update. Note also that Eqs.~\ref{uphi1} and~\ref{uphi2} are coupled with each other as well as with the parameters $\gamma_{p,g}^t$, $\gamma_{q,h}^t$. 

For the variational parameters $\Sigma_p^t$, we have for the diagonal components ($\sigma_{p,1}^t, \sigma_{p,2}^t, ...\sigma_{p,k}^t $): 
\begin{eqnarray}
{\eta_k^2 \over { \sigma_{p,k}^t}} =   1+(1-\beta_p)^2 + \sum_q  Y_t(p,q)\beta_q^2 {w_{q\leftarrow p}^t}^2\nonumber~~~~~~~~~~~~~ \\
 +2 \eta_k^2 (N-1)     {\sigma_{p,k}^t\over \zeta_{p}^t}  \exp(\gamma_{p,k}^t + {(\sigma^t_{p,k})^2 \over 2}),
 \label{sigma}
\end{eqnarray}
where $\eta_k$ is the diagonal component of the covariance matrix $\Sigma_\mu$. Similarly, we obtain equations  for  the variational parameters $\gamma$-s. Generally, those equations are different for $\gamma_{p,g}^0$, $\gamma_{p,g}^T$, and $\gamma_{p,g}^t$, $0< t < T$. Since those equations are too cumbersome, here we simply note that their general form is:
\begin{eqnarray}
\vec \gamma_{p}^t = f(\vec \gamma_{p}^{t-1},\vec \gamma_{p}^{t+1},\vec \gamma_{q}^{t}, \vec{\phi}^t_{p \rightarrow q},\vec{\phi}^t_{q \leftarrow p} ,\zeta_p^t, \Sigma_{p}^t),
\label{gamma}
\end{eqnarray}
Thus, the parameter $\vec{\gamma}_p^{t}$ depends on its past and future values,  $\vec{\gamma}_p^{t-1}$ and  $\vec{\gamma}_p^{t+1}$ , as well as the parameters of its neighbors.  Finally, for the variational parameters $\zeta$ we have  
\begin{equation}
 \zeta _{p}^t = \sum _i \exp(\gamma_{p,i}^t + {{\sigma^t _{p,i}}^2 \over 2} )
\end{equation}
Note that the above equations can be solved via simple iterative update as before. To expedite convergence, however, we combine the iterations with Newton--Raphson method, where we solve for individual parameters while keeping the others fixed, and then repeat this process until all the parameters have converged.

\subsection{Variational M step}
The M-step in the EM algorithm computes the parameters by maximizing the expected  log-likelihood  found in the E-step. The model parameters in our case  are: $B^t$, the role-compatibility matrix, the  covariance matrix $\Sigma_\mu$,  $\beta_p$ for each node, $w_{p\leftarrow q}^t$ for each pair, $\vec{\alpha}$, and $A$ from the prior. 
 
If we assume that the time variation of the block compatibility  matrix  is small compared to the evolution of the node attributes, we can neglect the time dependence in $B$, and use its average across time, which yields:  
\begin{equation}
\hat{B}(g,h) =    {{\sum_{p,q,t} Y_t(p,q) \cdot \phi ^t _{p \rightarrow q,g} \phi ^t_{p \leftarrow q,h}}\over {\sum_{p,q,t}  \phi ^t _{p \rightarrow q,g} \phi ^t_{p \leftarrow q,h}}} 
\end{equation}
 Likewise, for the update of diagonal components of the noise covariance matrix $\Sigma_\mu$, 
\begin{equation}
\hat{\eta}_{k} = {1\over N(T-1)}E_q[\sum_{p,t} (\mu_{p,k}^t - (1-\beta)\mu_{p,k}^{t-1} - \beta \mu_{\mathcal{S}(p,t-1) ,k} )^2 ]
\end{equation}
Similar equations are obtained for $\beta_p$ and $w_{p\leftarrow q}^t$. The update equation of $\beta_p$ and $w_{p\leftarrow q}^t$ is a function of $\gamma$ and $\sigma$ which are related to the transition for specific node $p$. Since these equations are rather involved, they will be provided elsewhere. 

The priors of the model can be expressed in closed form as below: 
\begin{eqnarray}
\vec{\alpha}^0 
 = {1\over N} \sum_p \vec{\gamma_p^0}
 \end{eqnarray}
\begin{eqnarray}
a_{k} = \sqrt{{1\over N} \sum(\gamma_{p,k}^2 + \sigma_{p,k}^2 -2\alpha_k^0 \gamma_{p,k} + {\alpha_k^0}^2)}
\end{eqnarray}

\section{Results}

\subsection{Experiments on Synthetic Data }
We tested our model by generating a sequence of networks according to the  process described above, for   $50$ nodes, and $K=3$ latent roles across $T=8$ time steps. We use a covariance matrix of $A = 3I$, and mean $\vec{\alpha}^0$ having homogeneous values for the prior, so that initially nodes have a well defined role (i.e., the membership vector is peaked around a single role). More precisely, the majority of nodes had around $90 \%$ of membership probability mass centered at a specific role, and on average a third of those nodes will have $90 \%$ on role $k$. For the role-compatibility matrix, we gave high weight at the diagonal.  

\begin{figure}[!t]
\centering
\subfigure[]{
    \includegraphics[width = 0.26\textwidth]{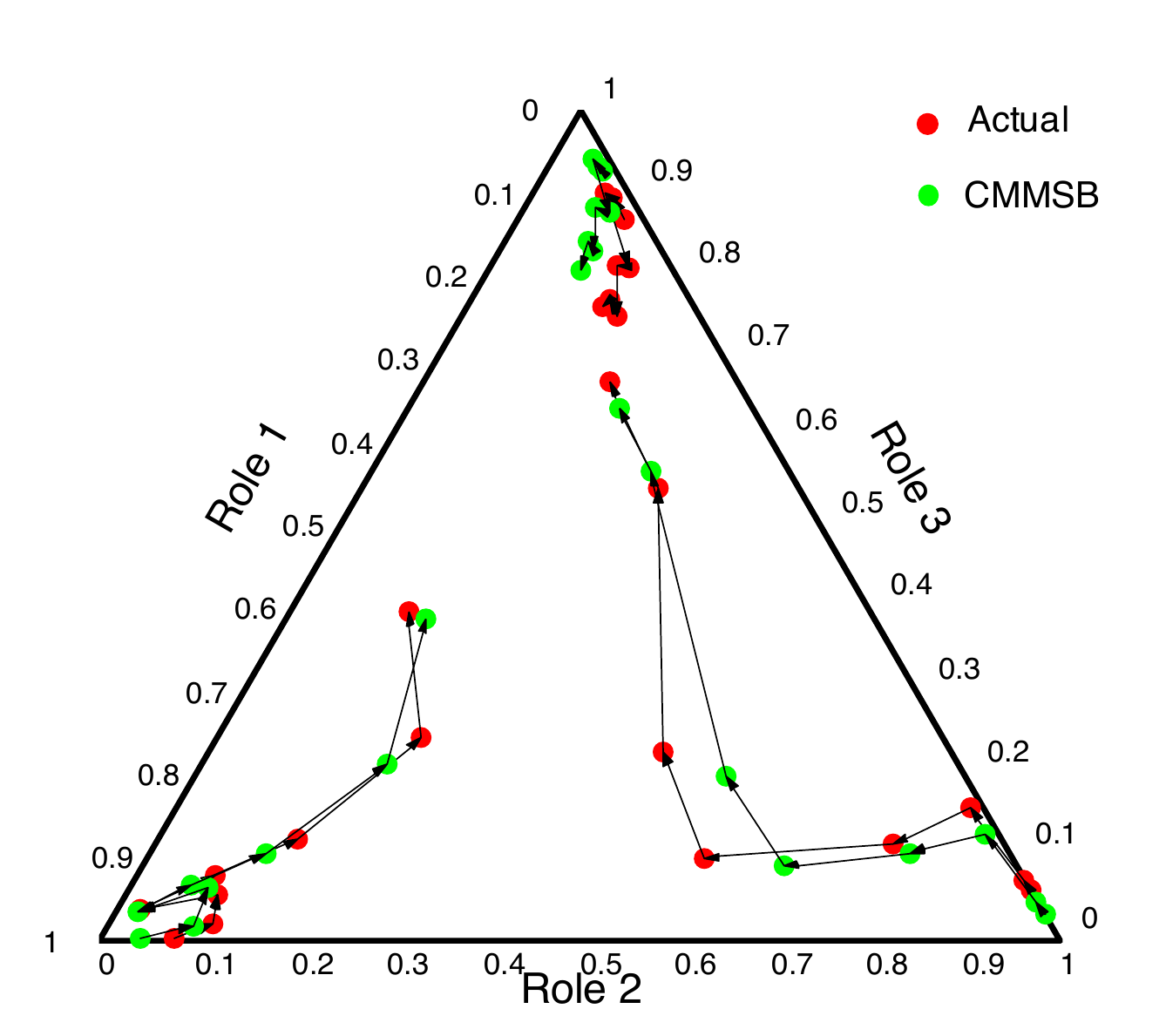} \label{fig:simplex}
    }
    \subfigure[]{
    \includegraphics[width = 0.27\textwidth]{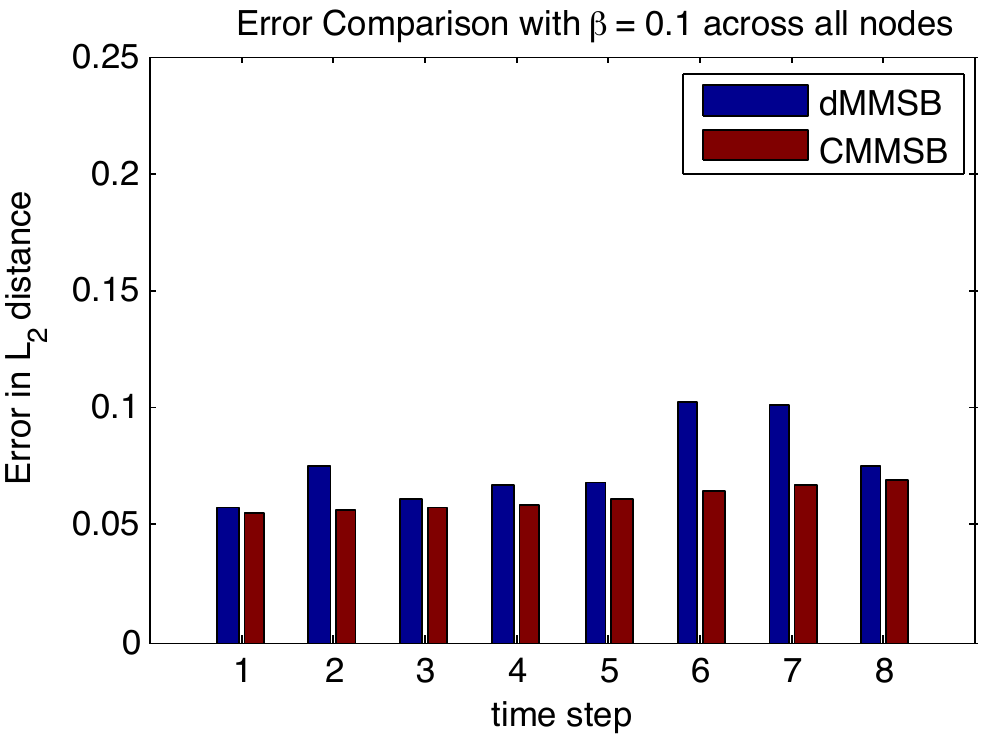} \label{fig:errorplot1}
    }
    \subfigure[]{
    \includegraphics[width = 0.27\textwidth]{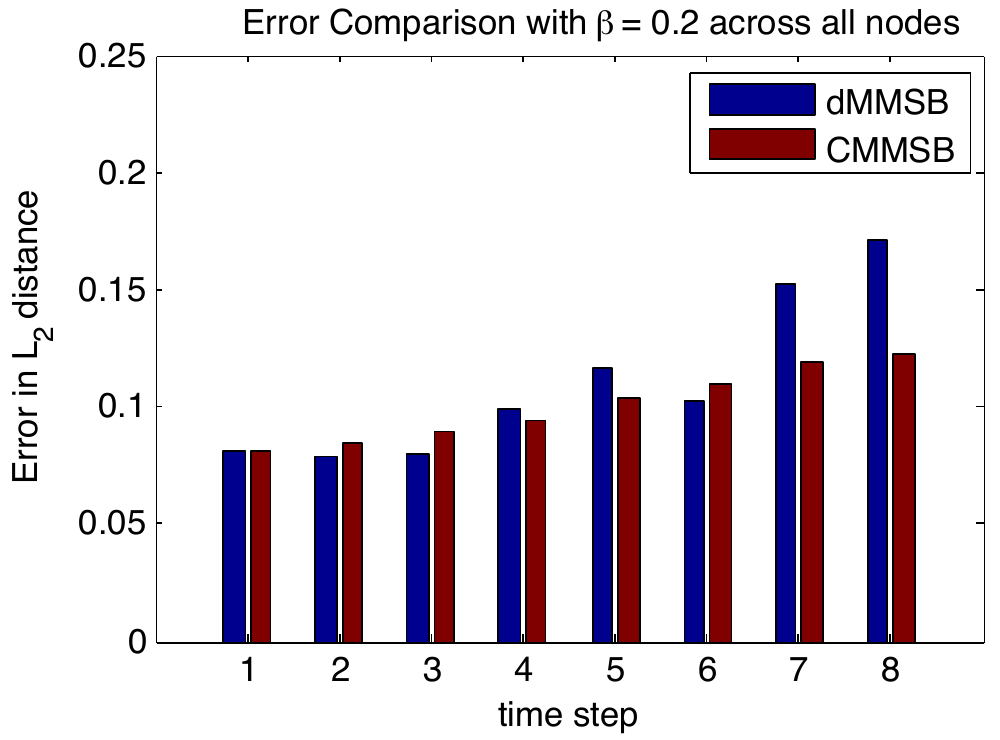} \label{fig:errorplot2}  
    }   
         \caption{(a) Actual and inferred mixed membership trajectories on a simplex. (b) Inference error for dMMSB and CMMSB for synthetic data generated with $K = 2$ and $\beta = 0.1$ for all the nodes (c) when $\beta = 0.2$ for all the nodes }
\end{figure}

Starting from some initial parameter estimates, we performed variational EM and obtained re--estimated parameters which were very close to the original values (ground truth). With those learned parameters, we inferred the hidden trajectory of agents as given by their mixed membership vector for each time step. The results are shown in Fig~\ref{fig:simplex}, where, for three nodes,  we plot the projection of trajectories onto the simplex.  One can see that for all three nodes, the inferred trajectories are very close to the actual ones. 

\subsection{Comparison with dMMSB}

As a further verification of our results, we compare the performance of our inference method to the dynamic mixed membership stochastic blockmodel (dMMSB)\cite{dMMSB}.
We use synthetic data generated in a manner similar to the previous section. This time, though, for simplicity we keep $K=2$ and we set all the $\beta$'s to some constant for all the nodes, $\beta = 0.1$ in one trial and $\beta = 0.2$ in the other. 
In this case, we compare performance by evaluating the distance in $L_2$ norm between actual and inferred mixed membership vectors for each method. At each time step, we calculate the average over all nodes of the $L_2$ distance from the actual membership vector. 

As shown in Fig. \ref{fig:errorplot1} and \ref{fig:errorplot2},  CMMSB captures the dynamics better than the dMMSB. This is due to the fact that our model tracks all of the nodes individually (internal dynamics), while dMMSB regards the dynamism as an evolution of the environment (external dynamics). 
Here, we have only included results for relatively small and homogeneous dynamics.
In fact, we noticed that 
our method tends to fare even better as we increase the degree of dynamics or the heterogeneity of dynamics across nodes (node-varying values of $\beta$). We believe heterogeneous dynamics is more prevalent in real systems, and so we expect our method to outperform dMMSB even more than is indicated by Fig.\ref{fig:errorplot2}.

\subsection{US Senate Co-Sponsorship Network}
We have also performed some preliminary experiments for testing our model against real--world data. In particular, we used senate co--sponsorship networks from the $97$th to the $104$th senate, by considering each senate  as a separate time point in the dynamics.  There were 43 senators who remained part of the senate during this period. For any pair of senators $(p,q)$ in a given senate, we generated a directed link $p \rightarrow q$ if $p$ co-sponsored at least $3$ bills that $q$ originally sponsored. The threshold of $3$ bills was chosen to avoid having too dense of a network. With this data, we wanted to test (a) to what extent senators tend to follow others who share their political views (i.e., conservative vs. liberal) and (b) whether some  senators change their political creed more easily than others.  

The number of roles $K=2$ was chosen to reflect the mostly bi--polar nature of the US Senate. The susceptibility of senator $p$ to influence is measured by the corresponding parameter $\beta_p$,  which is learned using the EM algorithm. High $\beta$ means that a senator tends to change his/her role more easily. Likewise, the power of influence of senator $q$ on senator $p$ is measured by the parameter $w_{p\leftarrow q}^t$, where $w_{p\leftarrow {q_1}}^t > w_{p\leftarrow {q_2}}^t$ means senator $q_1$ is more influential on senator $p$ than senator $q_2$. Here the direction of the arrow reflects the direction of the influence which is opposite to the direction of link. To initialize the EM procedure, we assigned the same $\beta$, and $w$ to all the senators, and start with a matrix which is weighted at the diagonal for $B$. 

Another method for validation is to compare the degree of influence. Our model handles, and learns, the degree of influence in the update equation. Sorting out influential senators is an area of active research. Recently, KNOWLEGIS has been ranking US senators based on various criteria, including influence, since 2005. Since our data was  extracted from the 97th senate to the 104th senate, direct comparison of the rankings was impossible. Another study\cite{maine} ranked the 10 most influential senators in both parties who have been elected since 1955. We compared our top 5 influential senators, and we were able to find 3 senators (Sen. Byrd, Sen. Thurmond, and Sen. Dole) in the list. 

\subsection{Interpreting Results}

The role-compatibility matrix  learned from the Variational EM has 
high values on the diagonal confirming our intuition that interaction is indeed more likely between senators that share the same role. Furthermore, the learned values of $\beta$ showed that senators varied in their ``susceptibility". In particular, Sen. Arlen Spector was found to be the {\em most influenceable} one,  while Sen.  Dole was found to be one of the most {\em inert} ones. Note that while there are no direct ways of estimating the ``dynamism" of senators, our results seem to agree with our intuition about both senators (e.g., Sen. Spector switched  parties in 2009 while Dole became his party's candidate for President in 1996). 

To get some independent verification, we compared our results to the yearly ratings that ACU (American Conservative Union), and ADA(Americans for Democratic Action) assign to senators~\footnote{Accessible at http://www.conservative.org/,\\ http://www.adaction.org/ }. ACU/ADA rated every senator based on selected votes  which they believed to have a clear ideological distinction, so that high scores in ACU mean that they are truly conservative, while lower score in ACU suggests they are liberal, and for ADA vice versa. 
To compare the rating with our predictions (given by the membership vector)  we scaled the former to get scores in the range $[0,1]$. 

\begin{figure}[h!] 
  \centering
    \includegraphics[width=0.25\textwidth]{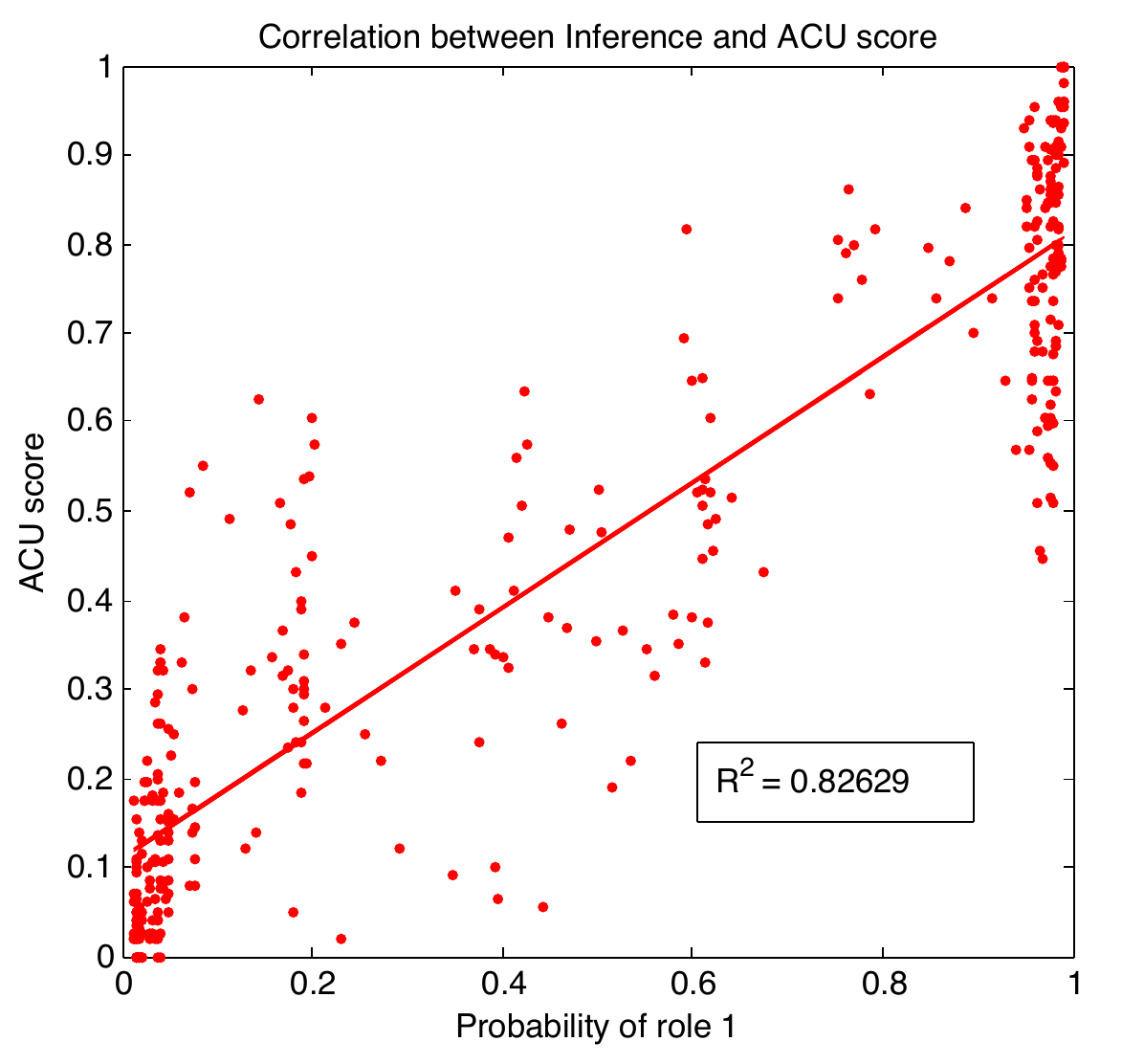}\\
    \includegraphics[width=0.25\textwidth]{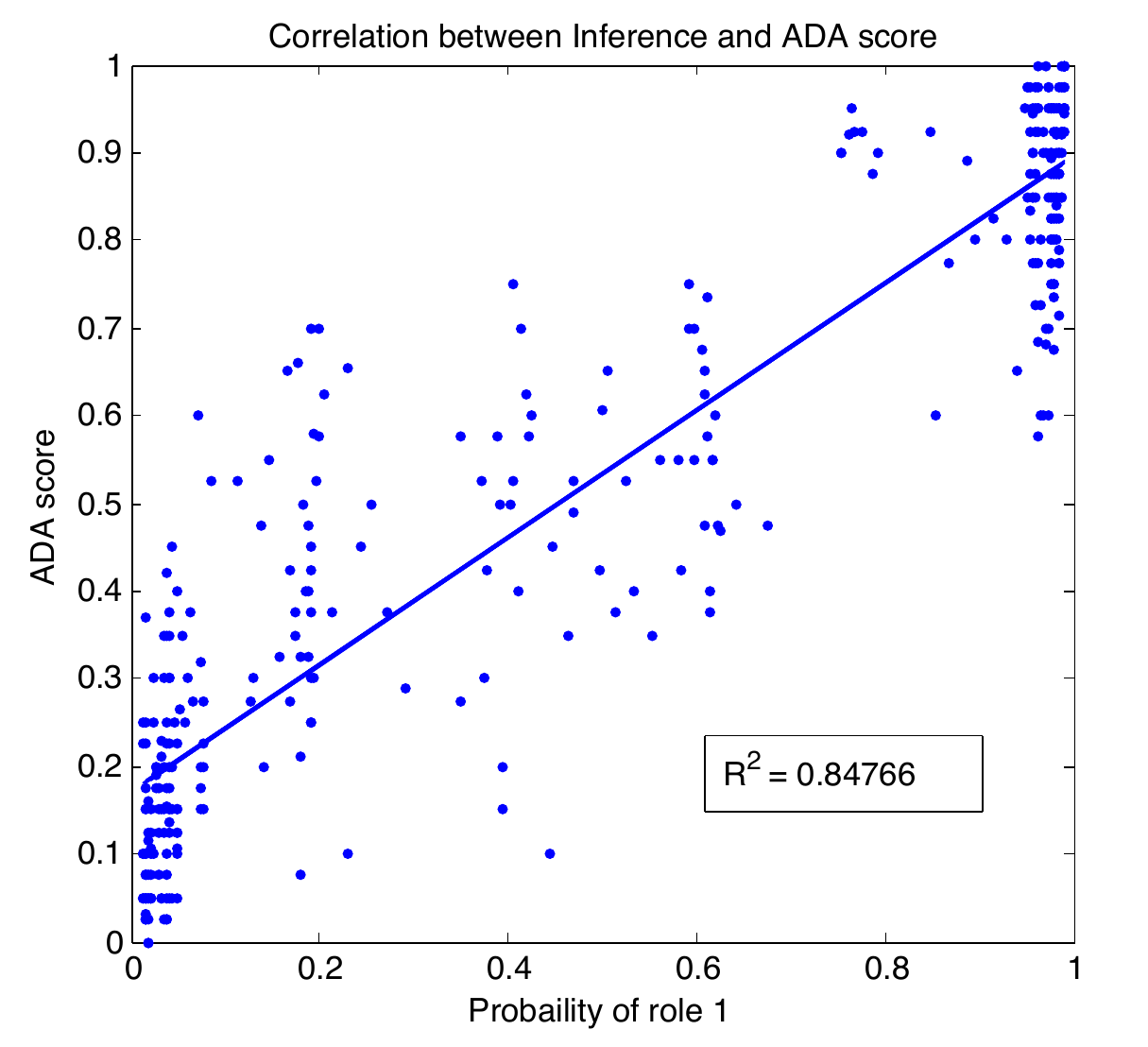}
\caption{Correlation between ACU/ADA scores and inferred probabilities.}
\label{correlation}
\end{figure}

Fig. \ref{correlation} shows the relationship between these scores and our mixed membership vector score, confirming our interpretation of the two roles in our model as corresponding to liberal/conservative. 
Although  those values cannot be used for quantitative agreement, we found that at least qualitatively,  the inferred trajectories agree reasonably well  with the ACU/ADA ratings. This agreement is rather remarkable since the ACU/ADA scores are based on selected votes rather than  co--sponsorship network as in our data.


\begin{figure}[h!] 
  \centering
    \includegraphics[width=0.26\textwidth]{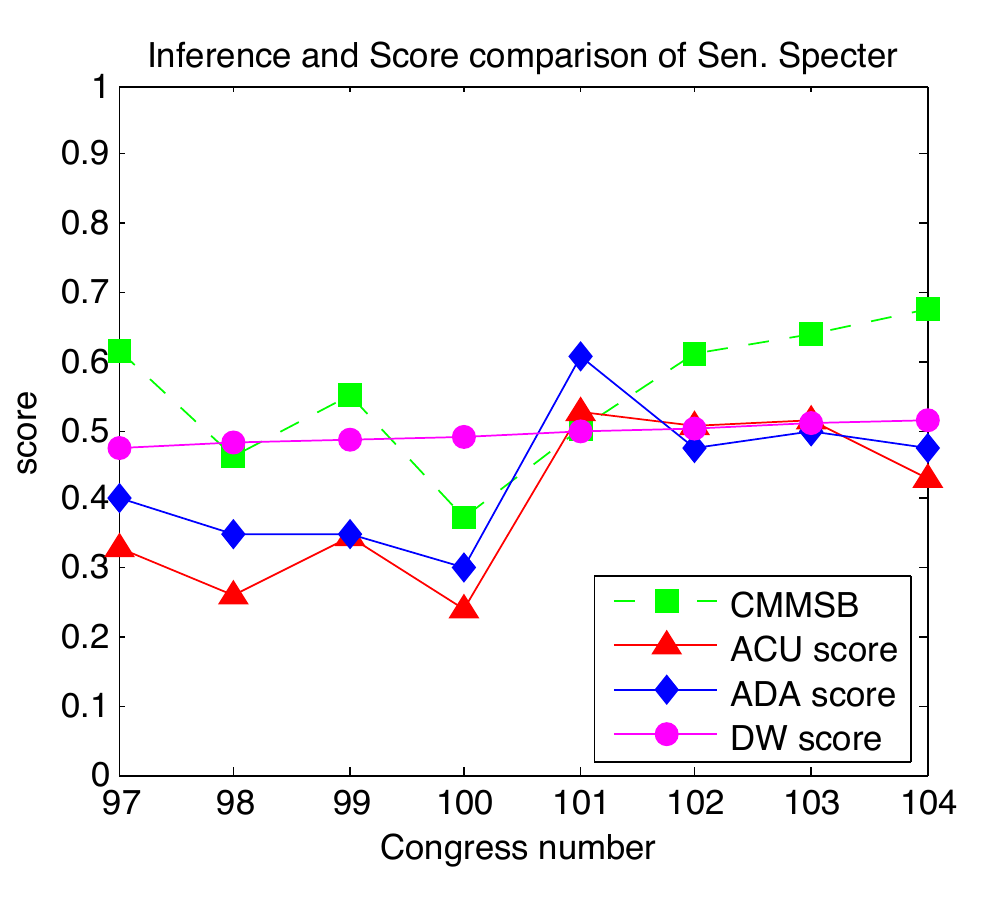}\\
    \includegraphics[width=0.26\textwidth]{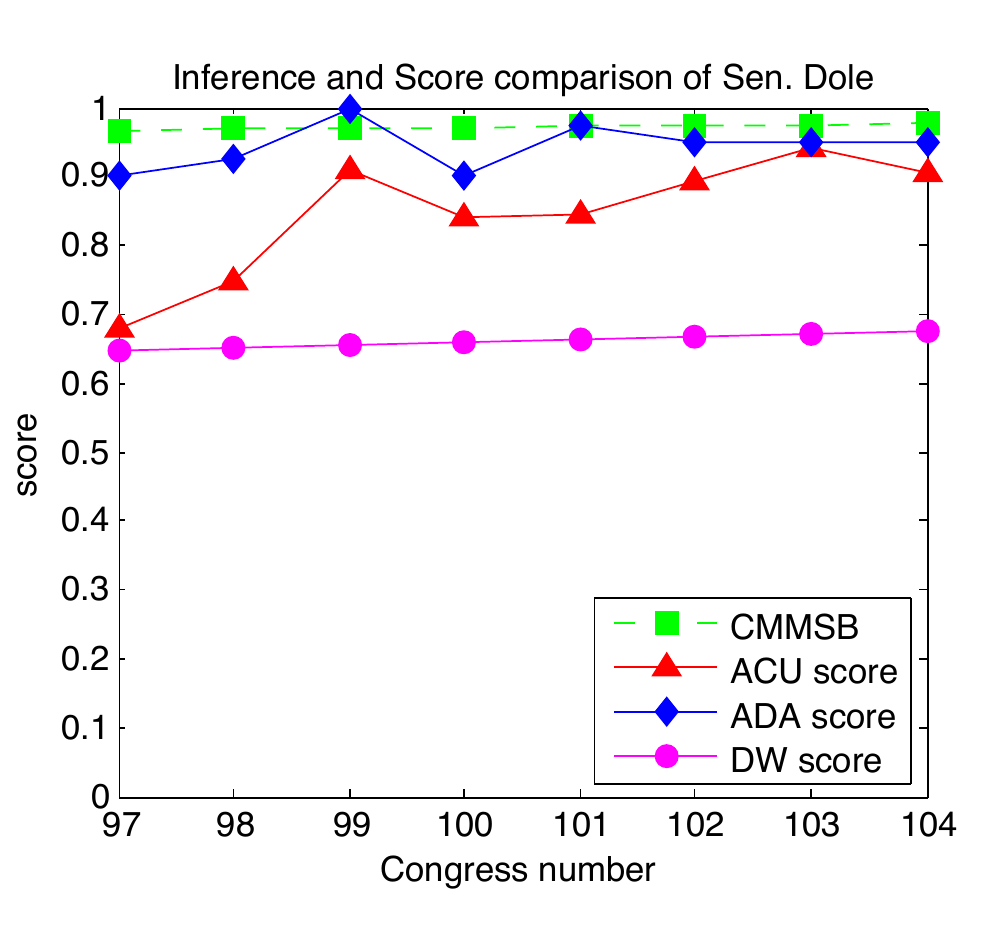}
\caption{Comparison of inference results with ACU and ADA scores: Sen. Specter (top) and Sen. Dole (bottom).}
\label{senator}
\end{figure}

Of course, we are most interested in correctly identifying the dynamics for each senator. We compare our inferred trajectory of the most dynamic senator, and the inert senator to the scores of ACU, and ADA. In Fig.\ref{senator} the scores of ADA have been flipped, so that we can compare all of the scores in the same measurement. However, since ACU/ADA scores are rated for every senator each year, the dynamics of inference, and the dynamics of ACU/ADA scores cannot be compared one to one. Not all senators showed high correlation of the trend like senator Specter, and Dole. 


\subsection{Polarization Dynamics}

 \begin{figure}[h!] 
  \centering
    \includegraphics[width=0.27\textwidth]{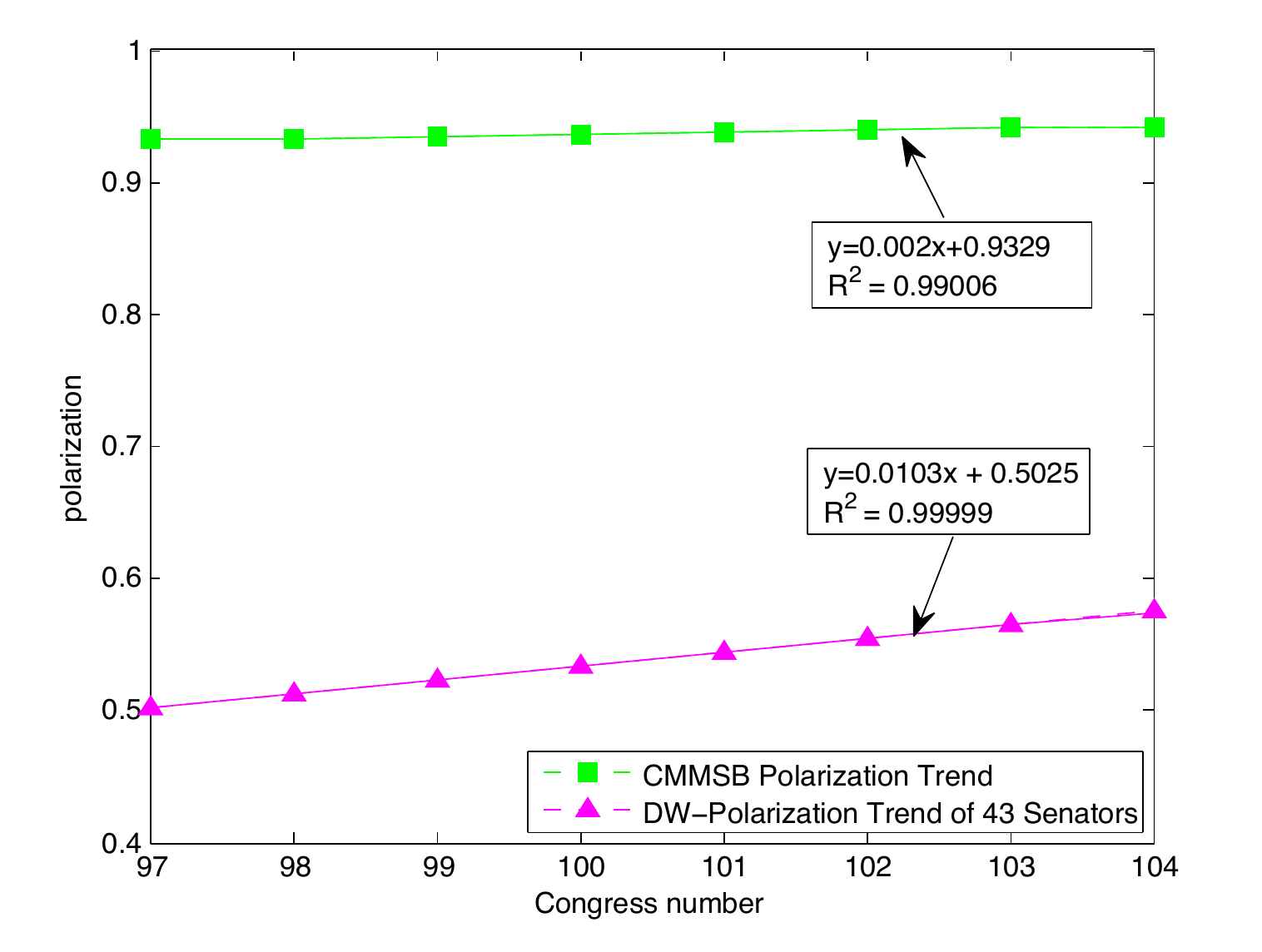}
\caption{Polarization trends during 97th--104th US Congresses.}
\label{tcc}
\end{figure}
The yearly ACU/ADA scores give a good comparison of the relative political position of senators scored in each year. However, they are not very appropriate for comparison between years, a point illustrated by the fact that the score is based on voting records for different bills in each year. Therefore, for validation of the dynamics we turn to another scoring system highly regarded by political scientists and used to observe historical trends, the DW-NOMINATE score. For the time period of our study, \cite{PA} shows that the political polarization of the senate was increasing. In particular, they show that the gap between the average DW-NOMINATE score of Republicans and Democrats is monotonically increasing, as we show in Fig. \ref{tcc}. In fact, the polarization for the entire senate was stronger every year. This is due to the unbalanced seats in the entire senate. In other words, our data had 22 Republican, and 21 Democratic, while for the entire senate, majority out numbered minority by around 10 seats. For comparison, for each time step we took the average of our inferred score for the 14 most and least conservative senators. As we show in Fig. \ref{tcc}, our inferred result agrees qualitatively with the results of \cite{PA}, showing an increase in polarization for every senate in the studied time-window. Since the DW-NOMINATE scores uses its own metric, and our polarization is measured by the difference between upper average and lower average probability, we should not expect to get quantitative agreement. We would like to highlight, however, that the direction of the trend is correctly predicted for each of the eight terms. 


\section{Conclusion}

We have presented the Co--evolving  Mixed Membership Blockmodel for modeling inter--coupled node and link dynamics in networks.  We used a variational EM approach for learning and inference with CMMSB, and were able to reproduce the hidden dynamics for synthetically generated data, both qualitatively and  quantitatively. We also tested our model using the US Senate bill co--sponsorship data, and obtained reasonable results in our experiments. In particular, CMMSB was able to detect increasing polarization in the Senate as reported by other sources that analyze individual voting records of the senators.  As a future work, we intend to test our model against different real--world data, such as co--authorship network of publications.  We also plan to extend  CMMSB in several ways. For instance, a bottleneck of the current model is that it explicitly considers links between all the pairs of nodes, resulting in a quadratic complexity in the network size. Most real world networks, however, are sparse, which is not accounted for in the current approach. Introducing sparsity into the model would greatly enhance its efficiency. 

\section{Acknowledgments}
This research was supported in part by the National Science Foundation
under grant No. 0916534, U.S. ARO MURI grant No. W911NFÐ06Ð1Ð0094, and US AFOSR MURI grant No. FA9550-10-1-0569.

\bibliographystyle{aaai}
\bibliography{Bibliography-File}

\end{document}